# From Microwave Anisotropies to Cosmology

Douglas Scott, Joseph Silk & Martin White

Center for Particle Astrophysics,
and Departments of Astronomy and Physics,
University of California, Berkeley, CA 94720-7304

**Fluctuations in the temperature of the cosmic microwave background have now been detected over a wide range of angular scales, and a consistent picture seems to be emerging. This article describes some of the implications for cosmology. Analysis of all the published detections suggests the existence of a peak on degree scales of height 2.4 to 10 (90%CL) relative to the amplitude of the power spectrum at large angular scales. This result confirms an early prediction, implies that the universe did in fact recombine, and limits theories of structure formation. Illustrative examples are provided of how the comparison of microwave background and large−scale structure data will be a potentially powerful means of answering fundamental questions about the universe.**

All indications are that the large−scale structure in the Universe has developed by the process of gravitational instability from small primordial fluctuations in energy density generated in the early universe. Slightly overdense regions collapsed under their own gravity to become more and more overdense, the density contrast increasing with time. In the 'standard' model, the fluctuations were laid down during an inflationary phase, when quantum fluctuations, inescapably present in any theory, were boosted in scale. The outcome of the growth of these fluctuations is the formation of galaxies and galaxy clusters by the present epoch.

There are three distinct approaches to studying the primordial fluctuations from which large−scale structure originated. Numerical simulations of non−linear growth and collapse have provided realistic descriptions of galaxies and galaxy clusters, and constrain the fluctuation spectrum over scales of a few Mpc in the recent past. On large scales, redshift surveys measure the density fluctuations directly, out to about 100 Mpc. On still larger scales, the COsmic Background Explorer (COBE) satellite and a series of subsequent experiments have detected temperature fluctuations in the cosmic microwave background. These temperature fluctuations are the high redshift precursors, emanating from a redshift of about 1000, of the fluctuations that generated the structures we see today.

Thus cosmic microwave background (CMB) fluctuations provide a key to understanding the origin of large-scale structure in the Universe. The ripples, or 'anisotropies', in the background radiation represent not only the initial seeds from which structure first emerged, but also contain coded measures of various cosmological parameters. Simultaneously, we are probing the thermal history of the early Universe. One consequence of the analysis described below is that very early reionization, popular in several cosmological models wherein structure forms early, cannot have occurred.

We are now on the verge of a measurement of the total density of the Universe, and there is a possibility of learning about any epoch of inflation from the detailed shape of the CMB anisotropy spectrum on large angular scales. On degree scales, we study CMB temperature fluctuations generated at the epoch of last scattering of the radiation. The large−scale structure (LSS) of galaxies in the Universe provides an independent measure of density fluctuations of similar physical size, but at the present epoch. By probing the fluctuations at two different times, the comparison of CMB and LSS measurements constrains the growth of fluctuations, which in turn depends on the total matter content of the Universe together with the value of the cosmological constant, $\Lambda$.

In the 'standard' (Cold Dark Matter, CDM, and its variants) model, one assumes that primordial gravitational potential fluctuations are generated in the inflationary era, and are visible on the last scattering surface as photons propagate out of the potential wells that are destined to eventually form clusters and superclusters of galaxies. The potential fluctuations drive photon density and velocity fluctuations, which lead to anisotropies in the observed temperature of the microwave sky. A schematic list of the sources of fluctuations (in rough order of importance with decreasing angular scale) is given in Table 1.

The largest scale anisotropy is the dipole generated by the motion of the Sun and Earth through the microwave background. The other effects arise as the photons interact with perturbations in the matter. For example the Doppler shift arises when the photons gain energy by scattering off moving electrons. For further discussion of all of these effects see (*1*). In this article we shall focus on the gravitational potential and adiabatic contributions, which are of primary importance on large and intermediate angular scales ($\gtrsim 5'$) in $\Omega_0 = 1$ cosmological models. (Here $\Omega_0$ is the total density of the universe in units of the critical density $\rho_c = 3H_0^2/(8\pi G)$ where $H_0$ is the Hubble constant today). The Doppler shifts are of importance only in scenarios where the universe was reionized by energy injection at late times, which we shall argue are now disfavored.

A frenzy of experimental activity followed the COBE DMR announcement in 1992 of the detection of fluctuations

(2). There have subsequently been no fewer than 15 claimed detections in different regions of the sky by some 9 independent experiments, 4 balloon and 5 ground-based. Most experiments have degree-scale resolution, although two detections almost overlap with the COBE scale of ~7 degrees. We present these detections in Fig. 1 and Table 2, where $Q_{\rm flat}$ is a measure of '$\Delta T$' to be precisely defined later.

Now that the existence of CMB fluctuations over a wide range of angular scales has been established (1), emphasis is shifting toward studies that try to fix the parameters of theoretical models. There have been several papers that combine the data from two experiments, usually the COBE DMR results on the largest scales plus a specific smaller angular-scale experiment, to place constraints on some cosmological parameters or models. We believe that it is now feasible to combine the available data from *all* of the high quality, generally multi-frequency, experimental measurements of microwave background anisotropies, to go one step beyond simply constraining the normalization. However, rather than trying to rule out specific cosmological models, we adopt a more phenomenological approach.

Firstly, we set up a 'toy-model' for the radiation power spectrum, which is flat on large angular scales and has a peak in power around multipole $\ell \simeq 250$, or scales of $0.^\circ 5$ (see Fig. 2). Secondly, we take the data from the different experiments and convert each of them into *one* measure of power, so that they can all be plotted together for comparison, and so that they can be combined to place constraints. Finally, we calculate the best-fitting height for the peak in our phenomenological power spectrum.

Despite the apparent scatter of reported $\Delta T/T$ values, we find that a distinct pattern is emerging. In particular a totally flat scale-invariant power spectrum is ruled out by the data (at the 99% confidence level), which instead prefer some sort of peak with height $\sim$ 4-5 relative to the large-angle part of the radiation power spectrum. This result is remarkably close to what theorists had been anticipating, and has interesting implications for cosmology.

Of course, we are assuming that the error estimates for individual experiments allow for possible foreground contamination: this we believe to be the case for most of the experiments utilized in our analysis. While there is still some concern that not all of the measurements see only CMB fluctuations, for several experiments the case against foreground contamination is quite compelling. We will proceed under the assumption that the data can be taken at face value and investigate what they appear to be telling us. We caution that the true cosmic fluctuation signal may be overestimated if contaminated by foreground, and underestimated if too much foreground has been subtracted: either of these effects may be present in some of the experiments. Particularly worrying are experiments with very limited frequency coverage or data sets with obvious 'contamination'. However with many data points contributing to our analysis, a 'wrong' experiment should not skew our conclusions unduly. Clearly, as the data improve and issues related to foregrounds are further understood, our conclusions can be refined. Nevertheless we are confident that our analysis should give a flavor of the kind of information already available from CMB studies, and an indication of what will soon be possible.

## The Radiation Power Spectrum

It is standard practice in CMB studies to work in terms of the multipole moments of the temperature anisotropy. One conventionally defines $C_\ell = \langle |a_{\ell m}|^2 \rangle$, where $\Delta T/T(\theta,\phi) = \sum_{\ell m} a_{\ell m} Y_{\ell m}(\theta,\phi)$, the angled brackets representing an average over the ensemble of possible fluctuations, and where $Y_{\ell m}$ are the spherical harmonics and $\theta$ and $\phi$ are angular coordinates on the sky. Assuming the fluctuations have a gaussian probability distribution, the models are uniquely specified by giving their $C_\ell$'s, which in any model are simply a function of the cosmological parameters.

Given an input primordial fluctuation spectrum, one can follow the distribution of photons, baryons, neutrinos and dark matter as the universe evolves. The output is the spectrum of anisotropies observable today. What is usually plotted is $\ell(\ell+1)C_\ell$ vs. $\ell$, which is the power per logarithmic interval in $\ell$, or a 2D power spectrum on the sphere. As an example, the solid line in Fig. 2 shows the anisotropy spectrum for the standard CDM model ($\Omega_0 = 1$, $\Omega_B = 0.05$ and $h = 0.5$, where the Hubble constant is $H_0 = 100\,h\,{\rm km\,s^{-1}\,Mpc^{-1}}$).

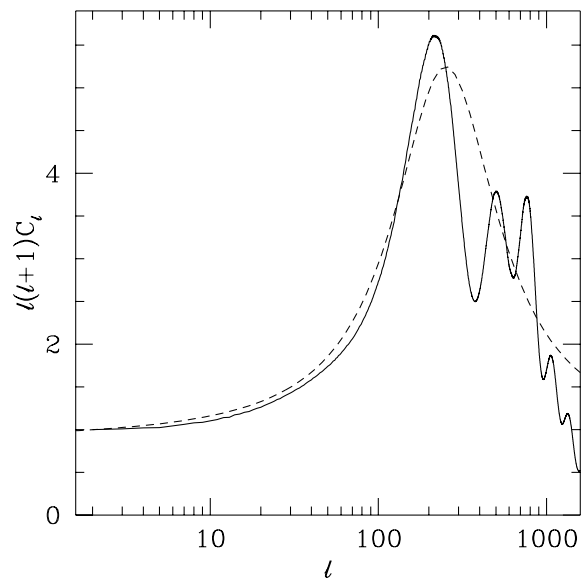

**Fig. 2.** The solid line shows the spectrum of temperature fluctuations for a standard Cold Dark Matter model. The quantity $\ell(\ell+1)C_\ell$ is power per logarithmic interval in multipole number $\ell$. Note that the curve is fairly flat at small $\ell$ (large angular scales) and has considerable structure at larger $\ell$ (small angular scales). Our simple phenomenological fit is shown by the dashed line. Note that the y-axis here is a measure of power, which is proportional to $Q_{\rm flat}^2$ (see Eq. 2).

In the spectrum, the large-angle (small $\ell$) 'plateau' is due to gravitational potential and large-scale, adiabatic, density perturbations, which are sensitive primarily to the dark matter fluctuations. A generic prediction of inflationary models is a primordial spectrum of adiabatic, density fluctuations with power spectrum $P_{\rm matter}(k) \propto k^n$ and $n \simeq 1$. For an $n \equiv 1$, or scale-invariant, spectrum, the temperature anisotropies are independent of angle at large angular scales, i.e. they have a flat angular power spectrum at small $\ell$. On the surface of last scattering, sound waves in the coupled baryon-photon fluid enhance (over the large angle value) the radiation power spectrum on scales around the horizon size at that epoch. (In the standard model, with reionization at redshift $\simeq 1000$, this effect sets in on scales below $\sim 1°$ or $\ell \sim 100$.) On the smallest scales the fluctuations are suppressed by photon diffusion (3,4,5,6) below $\simeq 5$ arcminutes ($\ell \simeq 1500$), the angle subtended by the width of the last scattering surface. Between the horizon and damping scales several peaks of successively smaller amplitude are generated. By measuring $\Delta T/T$ on various angular scales, it is possible to differentiate between the various contributions, and to begin the task of confronting the model predictions with the data in detail.

It has become apparent (7,6,8) that variations caused by different cosmological parameters are not 'orthogonal', in the sense that somewhat similar sets of $C_\ell$'s can be found for different parameters. Attempts to extract cosmological parameters from CMB data are further complicated by the fact that theories only predict the expectation values of the $a_{\ell m}$'s for an ensemble of skies and not the $a_{\ell m}$ on *our* sky (hence the 'cosmic variance'). These difficulties could be largely overcome with a high sensitivity, high angular resolution map of a large fraction of the sky, which would come from another satellite mission. Until then it will prove almost impossible to separate 'inital conditions' (e.g. predictions of inflation) from the evolution induced dependence on cosmological parameters using CMB data alone. While we do not work in a vacuum, and these problems may be partially overcome by using other observations, at present the required observations are highly uncertain, as we describe below.

However the current CMB fluctuation data are already capable of tackling other important issues. In particular, we will address constraints that may be placed on the thermal history of the universe. Could the universe have been ionized sufficiently early (at $z \gtrsim 50$) that the primordial degree scale CMB fluctuations would have been erased? It is possible to formulate this question in a form that the data may already be able to answer, by resorting to some theoretical prejudice and falling back on some assumptions that have been common in previous studies. We will assume that the power spectrum of radiation fluctuations is at least phenomenologically similar to that obtained from models like the 'standard' Cold Dark Matter (CDM) model, although we do not need all the dark matter to be cold. Specifically, we assume that the power spectrum is flat, corresponding to $n = 1$ on the largest scales, that $\Omega_0 = 1$, and that the tensor–mode (9) contribution to the fluctuations is small compared to that of the scalar modes (i.e. $C_2^T \ll C_2^S$). These assumptions have the advantage of minimality, however they have also received some support from analyses of the COBE two-year data, which prefer a power spectrum that is either flat or weakly rising towards high $\ell$ (10). This is marginally inconsistent both with an appreciable tensor mode (and the associated 'negative' tilt: $n < 1$) or a low value of $\Omega_0$, in a spatially flat background (11,12), though current limits are not very strong. Under these assumptions the most prominent feature of the theoretical power spectrum is the rise near $\ell \sim 200$, and the series of peaks at larger $\ell$'s (see Fig. 2).

The bumps and wiggles at $\ell$'s of a few hundred in the radiation power spectrum (13,14,15), known as adiabatic peaks, are caused by sound waves propagating in the baryon–photon fluid before the Universe recombined. These wiggles would also be seen in the matter power spectrum if the Universe were dominated by baryons; the absence of oscillations in a model like CDM is because the dominant component of the matter is not coupled to the photons. As the universe evolves, perturbations on large scales are unstable to their own gravity, while on smaller scales perturbations oscillate as sound waves driven by gravity, with a restoring force from fluid pressure. The scale which is just large enough to collapse is known as the Jeans scale, which initially grows with time. The different peaks and troughs correspond to photon density and velocity perturbations which have had an integral number of half–oscillations before the Jeans scale reaches their size, with complications caused by the dark matter potential wells and the thickness of the last scattering surface. Higher $\Omega_B$ corresponds to fewer photons per baryon and thus less pressure. This leads to a smaller Jeans length at any epoch, allowing perturbations to grow more before the Jeans scale reaches their size and they start to oscillate. The oscillations will therefore be of greater amplitude for higher $\Omega_B$, leading to higher adiabatic peaks when the photons are last scattered. The exact heights of the various bumps and wiggles come from a combination of potentials (for the first peak) and density and velocity effects (for all the peaks) and so depend on the specifics of the cosmological model. For example the height of the first peak is fairly insensitive to $h$ when $\Omega_B \simeq 5\%$, while the relative heights of subsidiary peaks have quite a strong $h$ dependence. However, experiments are sensitive to a wide range of $\ell$, which will somewhat wash out these variations.

The position of the first adiabatic peak depends essentially only on the geometry of the Universe (16,17). Spatial curvature in an open universe causes light rays to diverge as they propagate from the last scattering surface to the observer. Thus a fixed length scale subtends a smaller angle in an open universe. Specifically the size of the horizon at last scattering subtends an angle corresponding to $\ell \simeq 220 \Omega_0^{-1/2}$, with a small amount of Hubble constant dependence. So for an $\Omega_0 = 1$ model (our assumption), the position of the first adiabatic peak is well-determined. An extremely significant step for the near future will be when the data are up to the task of testing these assumptions and obtaining a firm constraint on $\Omega_0$.

The damping scale of the $C_\ell$'s is also a fairly robust physical quantity, determined by the thickness of the last scattering surface. The damping comes from photon diffusion out of overdensities (and into underdensities) on scales equal to the mean free path of the photons times the duration of recombination. If the universe recombines at redshift $z \simeq 1000$ and $\Omega_0 = 1$ this scale is about 5 arcminutes. Reionization at late times generates a new last scattering surface at lower redshift, moving the damping scale to lower $\ell$. If the universe reionized at sufficiently high redshift ($z \gtrsim 100$), and remained ionized until the present, the damping is sufficient to remove the peaks on degree scales.

We find that the radiation power spectrum can be reasonably approximated by a constant power spectrum, plus a Lorentzian peak located at $\log_{10} \ell = 2.4$ of width $\log_{10} \ell = 0.38$. Analytically we take

$$\ell(\ell+1)C_\ell = 6C_2 \left\{ 1 + \frac{A_{\text{peak}}}{1+y(\ell)^2} \right\} \bigg/ \left\{ 1 + \frac{A_{\text{peak}}}{1+y(2)^2} \right\} \quad (1)$$

with

$$y(\ell) = \frac{\log_{10} \ell - 2.4}{0.38},$$

where the amplitude of the Lorentzian at $\ell = 2$ has been divided out so that $A_{\text{peak}}$ is the height of the peak above the low-$\ell$ plateau. This is plotted as the dashed line in Fig. 2.

The parameters for the center and width of the Lorentzian were fitted to accurate $C_\ell$'s for a standard CDM model, which is the solid line in the figure. Our fitting function has the virtue of simplicity, and although it will not be a good fit in the complicated adiabatic peaks region, it is a extremely good on the rise to the main peak, where most of the experimental data points lie (see *18* for further discussion). We also note that the detailed *shape* of the power spectrum rising into the peak is an important theoretical prediction which can be checked by future experiments.

## Different Experimental Results

In order to use the results from several experiments at once, we need to convert them into a consistent system. The most straightforward and robust datum from each experiment is the total measured power. A simple parameterization of this power, integrated across the window function (bandpass) of the experiment, is given by the amplitude of a *flat* power spectrum, $\ell(\ell+1)C_\ell = \text{constant} = (24\pi/5)\,(Q_{\text{flat}}/T_{\text{CMB}})^2$, required to reproduce the measured power:

$$\text{Power} = \frac{1}{4\pi}\sum_{\ell=2}^{\infty}\frac{24\pi}{5}\frac{(2\ell+1)}{\ell(\ell+1)}\left(\frac{Q_{\text{flat}}}{T_{\text{CMB}}}\right)^2 W_\ell. \quad (2)$$

Here $W_\ell$ is the window function of the experiment, which defines the sensitivity of the experiment to any given scale (*19*). The constants in this expression have been chosen so that $Q_{\text{flat}}$ has the same meaning as the familiar root-mean-square power spectrum estimated quadrupole $Q_{\text{rms-PS}}$ for $n=1$ (see also *20*).

Our estimated values for different experiments are shown in Fig. 1. Each point represents a fit for the amplitude of a flat spectrum convolved with the specific window function of the experiment. The vertical error bars are $1\sigma$ errors on this amplitude, while the horizontal lines show the widths of the window functions at half peak height (and so should not be regarded as error bars). For the error bars on the 'power', we have taken them to be symmetric in $Q_{\text{flat}}$, or the same quantity as a '$\Delta T/T$' measurement.

We have chosen only to use quoted detections (see Table 2), and to neglect experiments that have given upper limits. Generally the error bars on these upper limits are large enough that they would not affect our results (e.g. *21*). On Fig. 1, we have shown three of the tightest constraints at smaller angular scales. The upper limits are plotted as 95% CL error bars, for the White Dish (*22*), OVRO (*23*) and ATCA (*24*) experiments (see *25* for more details). [These upper limits may pose constraints for open or flat models with low-$\Omega_0$, but do not strongly constrain standard CDM models.]

We note that by performing a fit of the functional form of Eq. 2 to the CMB data to get each $Q_{\text{flat}}$, we automatically insure that both the cosmic and sample (*26*) variance are fully included in the error analysis. We also add the quoted calibration uncertainty in quadrature to all of the error bars. If the power spectrum was *actually* a pure $n=1$ power-law, then the points would scatter about a horizontal line on this plot. The fact that there would appear to be a trend for the smaller angular-scale experiments to lie above such a line will be examined next.

## The Height of the Adiabatic Peak

Taking the data from Table 2 and the toy-model power spectrum of Eq. 1, we can employ a likelihood analysis to fit the two parameters using the data, i.e. the overall normalization and the height of the adiabatic peak. For each set of parameters we use Eq. 1 and the window functions to 'predict' Q for each experiment. These are then compared with the data in Table 2. Contrary to common wisdom, we find that the best fit 'peak model' is allowed at the 60% CL, showing that there is no *statistical* reason to increase the error bars on the points. The fact that at least one model provides a reasonable fit to the data in Table 2 is a sign that the experimental situation on degree scales is more coherent now than it was even last year, and is indicative of the very rapid experimental progress in this field.

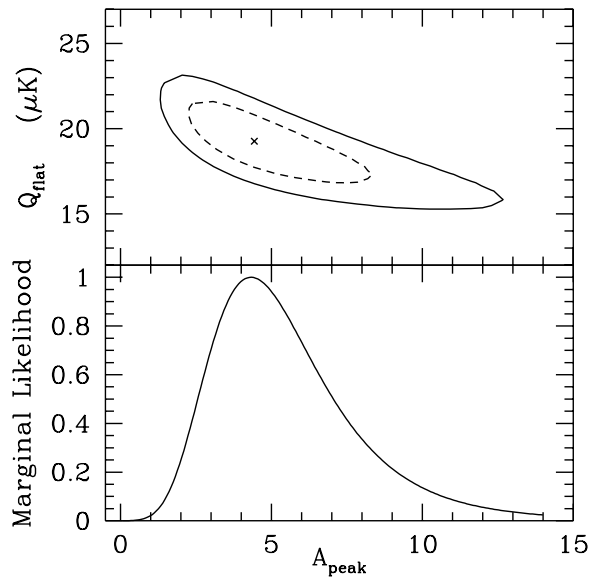

**Fig. 3.** (**A**) Contours of $\chi^2$ for a fit to the data of Table 2 using our Eq. 1. The cross marks the best fit ($Q_{\text{flat}} = 19\mu$K, $A_{\text{peak}} = 4.4$), while the contours mark 68% and 95% CL regions for the fit parameters. (**B**) The marginal likelihood, or likelihood integrated over $Q_{\text{flat}}$, as a function of $A_{\text{peak}}$ for our fitting form in Eq. 1. The likelihood has been normalized to peak at 1.

A plot of the 68% and 95% allowed regions in the parameters $Q_{\text{flat}}$ and $A_{\text{peak}}$ is shown in Fig. 3A. The power spectrum normalization is well fixed by large-scale measurements. To focus on the adiabatic peaks, we show (Fig. 3B) the 'marginal likelihood' or $\mathcal{L}(Q_{\text{flat}}, A_{\text{peak}})$ integrated over $Q_{\text{flat}}$ (assuming a uniform prior distribution for $Q_{\text{flat}}$). The best fit is $A_{\text{peak}} \simeq 4$, with mean $\simeq 5.5$, and $2.4 \leq A_{\text{peak}} \leq 10$ with 90% confidence (*27*).

## Cosmology with CMB Fluctuations

So what is all this telling us? For almost thirty years, there has been the promise of learning the answers to some truly fundamental questions by studying the anisotropies on the microwave sky. While experiments were giving only upper limits, the emphasis was on predicting the level of fluctuations from various cosmological theories. Now that fluctuations have been detected on a range of scales, theorists have been exploring in detail the predictions for the spectrum as a whole.

Our analysis suggests that there is an increase in the power measured on degree scales over that measured on larger scales. If we interpret this increase as due to an adiabatic peak (by far the most natural and compelling explanation), then there *were* oscillating baryon fluctuations in the early Universe, and early reionization could not have played a significant role in erasing primordial anisotropies in the CMB. The Thomson scattering optical depth since the Universe became ionized must have been small. This is the first definitive evidence that the Universe did in fact recombine (c.f. ref. *28*).

Seeded models generally are expected to have early non-linearity and early reionization. In calculations to date early reionization has been assumed, which suppresses the peaks at $\ell \sim 100$ (*29*). However, reionization is not *inevitable* in such models, and without it we would expect adiabatic peaks, although no explicit calculation has yet been done. Certainly the microwave background would be expected to be non-Gaussian on degree angular scales (roughly the horizon size at last scattering) in such texture or monopole models, though not in string models. The similarity in fluctuations in the three independent dust-free regions scanned by the MAX experiment may already be evidence against strongly non-Gaussian models.

The existence of the adiabatic peaks on degree scales would confirm a fundamental theoretical prediction (*30,3,13*), that complements the large angular scale COBE DMR detection of (presumably) gravitational potential fluctuations (*31*) predicted by inflationary cosmology. The latter are acausal reflections of the initial conditions at the end of inflation; the former provide a glimpse of the physics of the 'dark ages' of the Universe, long before the most distant galaxies or quasars had formed, via the possibility of probing back to the surface of last scattering.

The fact that our fitting formula, which has a plateau for low $\ell$, before rising into the peak, manages to pass through much of the data could also be taken as evidence against the Baryonic Dark Matter (BDM also called primordial isocurvature baryons: PBI) model, which rises rapidly into the adiabatic peaks and is not flat at large scales (*32,33*). This is just one example of the kind of information that accurate mapping of the adiabatic peaks will give. Another intriguing possibility may lie just ahead. So far, the data cannot accurately pin-point the position of the main peak, but it is clear that in the next few years this situation should improve. If it turns out that $\ell_{peak} \simeq 500$ rather than 200, then this will be very strong evidence that the Universe is open, since the open geometry makes the same physical scale subtend a smaller angular scale. Alternatively, should the data show a rapid rise near $\ell \simeq 200$ this would be hard to reconcile with open models.

Already, it seems that open models with $\Omega = 0.1$–$0.2$ have a hard time fitting both the large and degree scale data if there has been *no* reionization. Early reionization is at least as likely as in flat models, however, and due to the longer path length to a given redshift, the redshift for which the Universe becomes optically thick ($\tau \simeq 1$) is slightly reduced in open models (*34*). The shape of the power spectrum should be distinguishable from the $\Omega_0 = 1$ case, even with partial reionization. It seems that open models with $\Omega_0$ up to about 0.3 will be fully testable within only a few years.

## The CMB and LSS Together

While some qualitative features can already be derived from the CMB, what has become obvious is that it will be difficult to disentangle the variation due to simultaneous changes of different cosmological parameters until a high resolution, high sensitivity map of the sky can be obtained. This effect is sometimes known as 'cosmic confusion' (*7,6,8*). In particular, it seems that 'proving' inflation by simultaneously extracting $C_2^T/C_2^S$ and $n$ from the $C_\ell$'s will be very difficult (*35*), particularly because of cosmic variance at small $\ell$ (*36*). In the near future, it seems more likely that the questions being explored will be concerned with more 'classical' cosmology: $\Omega_0$, $\Omega_B$ and $h$. On the largest scales, there are already constraints on models which give the spectrum a negative slope, i.e. a combination of $C_2^T/C_2^S > 0$, $n < 1$ and cosmological constant ($\Lambda$CDM) or low–$\Omega_0$ inflationary models. The situation here will improve, particularly after the full four years of COBE data have been analysed. On smaller scales there is a wealth of cosmological information to be gained from the shape of the bumps and wiggles. However, it will be difficult to accurately separate the various effects without a major new, presumably space-based, experimental effort.

In order to break the degeneracy between the variations caused by different cosmological parameters, we can use constraints from other areas of astrophysics. A particularly fruitful area of study is the combination of CMB measurements with measurements of Large–Scale Structure (LSS). These two fields provide independent probes of the power spectrum in complementary regions (with some overlap), at different cosmological epochs. In Fig. 4 we show an example of the sort of information now available in these two fields (c.f. (*1*)): the matter power spectrum inferred from both CMB and LSS observations (the latter from (*37*)). It is important to realize that the conversion from the CMB to the matter power spectrum is very theory dependent; we have *assumed* a specific model in order to convert the numbers in Table 2 into the boxes shown on the plots. We show in the top panel of Fig. 4 both a standard CDM model and a mixed dark matter (MDM) model (i.e. CDM with a component of massive neutrinos). The middle panel shows an inflationary open universe CDM model chosen to satisfy Big Bang Nucleosynthesis, typical recent determinations of $H_0$ and the LSS *shape* constraint $\Omega_0 h \simeq 0.25$. The apparent divergence at small $k$ is an outcome of a specific inflationary model (*38,39,40*). The bottom panel shows a cosmological constant dominated model, with the same parameters as the open model, but $\Omega_\Lambda = 0.7$ to make the universe flat.

We see in the top panel that CDM normalized by the CMB predicts too much power on small scales, as is well known (*12*). An MDM model predicts less small–scale power, but perhaps too little to form galaxies early enough, and is also not a perfect fit to the shape. Tilting the model from scale invariance ($n < 1$), adding a tensor component, lowering the Hubble constant (*41*) or introducing decaying neutrinos (*42*) are possibilities for fixing both of these problems. However, the fit for the CMB alone is fairly good, as indicated by the scatter in the boxes both above and below the curves. Note also that the fraction of hot dark matter has negligible effect for degree-scale anisotropies.

In the middle panel we see that the inflationary open model manages to agree with the large angle CMB and LSS data quite well, but these models predict a falling $\ell(\ell+1)C_\ell$ on COBE scales which may be definitively tested in future. Also looking at the boxes, we see that it predicts fluctuations on degree scales which are somewhat small, though this is more a reflection of the parameters chosen (a low $\Omega_B$ due to the high $h$) than a generic prediction of open models.

In the bottom panel, the normalization inferred from CMB and large–scale structure observations has galaxies *anti*–biased, i.e. they need to be *less* clustered than the dark matter. This point has been realized before (*43,37*); however the increased CMB normalization has strengthened it (*44*). Notice that for the normalization of (*37*) the $\Lambda$ models predict too much power on LSS scales, despite the reduced growth rate due to lowering $\Omega_0$. However, one can lower the small scale power in *all* of these models by tilting the power spectrum away from $n = 1$ and/or attributing some of the large–angle CMB anisotropy to gravitational waves (*47,46*). Additionally there is still some freedom in the normalization of the matter power spectrum, through the biassing of galaxies relative to dark matter.

Now turn to Fig. 5. Here we have chosen different parameters for each of the classes of models from Fig. 4, to further demonstrate the power of using a range of CMB and LSS constraints together. In each case we have chosen somewhat 'non-standard' models, which are perhaps more realistic (e.g. by having some tilt: $n < 1$) as well as providing generally better fits to the data. The top panel shows a CDM model with $n = 0.9$, $h = 0.45$ and $\Omega_B h^2 = 0.02$, as may be suggested by the most recent nucleosynthesis considerations (*48*). The increase of $\Omega_B$ approximately counteracts the effect of lowering $n$ on the height of the first CMB peak. This model fits the data fairly well, with a reasonable level of power on cluster scales, and a passable fit to the shape around the turnover in the matter power spectrum. It would be possible to further decrease the power on small scales by allowing $C_2^T > 0$, but it becomes harder to accomodate an appreciable peak near $\ell \sim 200$ as $C_2^T$ is increased. By allowing similar parameter freedom one can also find MDM models which fit the data as well or better than the one in Fig. 4, but which are tilted and/or have gravity waves.

The open model in Fig. 5 has $\Omega_B h^2 = 0.02$ again, with $\Omega_0 = 0.4$, $h = 0.7$ and $n = 1$. Raising $\Omega_0$ and $\Omega_B$ provides a better fit to the degree scale CMB data, while lowering $h$ keeps the age, shape and small scale power roughly constant. The $\Lambda$-dominated model with $\Omega_0 = 0.3$, $h = 0.8$ and now $\Omega_B h^2 = 0.02$, has been tilted to $n = 0.95$, with a contribution of gravity waves $C_2^T = 0.35 C_2^S$. This is enough to stop the galaxies from being anti-biased as in Fig. 4. These two figures show that as the data improve it will be possible to considerably narrow the range of viable models.

## Other Parameters

With the ongoing explosion in the amount of useful data, it will soon be possible to simultaneously set some constraints on a number of different cosmological parameters. For the moment we have concentrated on a simpler question — obtaining *two* constraints from the data rather than just one. Already, the information we have gleaned has proved of interest for cosmology. In this section we look at some parameters that we could choose to constrain instead of $A_{\text{peak}}$.

In the context of an inflationary dark matter based theory, we can ask for information on the primordial spectral slope $n$. Determining this parameter accurately is well beyond the scope of this work, requiring a multi–parameter fit. However the 'peak' in the data at $\ell \simeq 200$, in combination with the COBE measurement, allows us to set a lower limit on $n$. Such a lower limit is most conservative if we ignore the possibility of gravity waves. From Big Bang Nucleosynthesis, we know that $\Omega_B$ cannot be arbitrarily large; in fact a value of $\Omega_B$ larger than 10% seems unlikely. Since the adiabatic peak height increases with $\Omega_B$, a lower limit on the tilt of such a model is a *conservative* lower limit for any model with a more reasonable value of $\Omega_B$. We find that the CMB data alone, even with such an unusually large $\Omega_B$, appear to require $n > 0.8$ at the 95% CL. This limit is competitive with combinations of large-angle CMB and LSS data (*49*). We should point out that for models based on inflation the spectral index is not expected to be *exactly* 1. Including some amount of tilt will be an important complication for future work, and will change the heights of the inferred peaks and the amplitude of fluctuations on smaller scales.

We can crudely convert our limit on $n$ into a limit on the optical depth, $\tau$, of the Universe (now once more assuming that $n = 1$). Recall that degree-scale anisotropies are reduced by $e^{-\tau}$ in a universe with significant reionization (*50*), and by $\ell^{(n-1)/2}$ in a tilted model. Thus our limit above, which compares the COBE scales ($\ell = 2$) to the adiabatic peak scales ($\ell \simeq 200$) translates into $\tau \lesssim 0.5$. If we assume $\Omega_B \leq 0.1$ and full ionization fraction ($x_e = 1$) from $z_{\text{ion}}$ until today, this corresponds to a redshift of reionization $z_{\text{ion}} \lesssim 50$: i.e. the Universe must have been neutral between redshifts 50 and 1000.

Knowledge of other cosmological parameters would to some extent affect our fits. For example, if $\Omega_0 < 1$, $\Lambda > 0$ or $C_2^T/C_2^S > 0$ etc., then the height of the adiabatic peaks will change relative to the COBE normalization. For example a combination of $n < 1$ and $C_2^T/C_2^S > 0$ would lower the predicted peak height relative to large scales, and may be preferred if the fit to $\Omega_B$ from Big Bang Nucleosynthesis is more than a few percent. Because of these possibilities it is hard to constrain $\Omega_B$ at present. However, very low $\Omega_B$, as inferred from recent primordial deuterium measurements, would be in conflict with an appreciable peak height.

The indications are that none of these complicating effects are so important as to invalidate our general result: the adiabatic peak is poking up above the noise. More ambitious analyses could clearly be done, but we feel the data do not yet warrant multi–parameter fits. In particular we have avoided the temptation to derive any specific cosmological quantity instead of our phenomenological amplitude $A_{\text{peak}}$. However, if we were to adopt a particular model, like standard CDM (with $h = 1/2$, say), then there obviously *is* a best–fitting baryon fraction. Recalling that our fitting form somewhat over–estimates power for models near the peak, our result, $A_{\text{peak}} \simeq 4$–6, would correspond to $\Omega_B \simeq 5\%$ for this specific model. At the moment this result is almost meaningless as a measurement of $\Omega_B$, since it depends sensitively on what is assumed for the other parameters. But as the CMB data and other astrophysical constraints improve, this technique is likely to complement the conventional BBN method, providing a measurement of $\Omega_B$ that has a very different $h$ dependence. Perhaps one day a combination of BBN and refined CMB measurements will constrain $h$!

## Conclusions

In summary, we believe that the new, intermediate angular scale CMB anisotropy data provide support for the existence of an adiabatic peak. This already has dramatic implications for the early Universe: recombination occurred on schedule, at a redshift of $\simeq 1000$ and the Universe remained neutral until a redshift less than $\simeq 50$. The primordial power spectrum is not too far from scale–invariant, and the increase in power on degree scales is *consistent* in position and amplitude with the adiabatic peak predicted by dark matter dominated models at the critical density. The former was already hinted at by the COBE DMR experiment in 1992, and subsequent experiments have provided some evidence for the latter. This development represents an important advance in our modelling of the deviations from uniformity in the early Universe, that complements, and potentially surpasses, our emerging understanding of the very early Universe origin of the primordial density fluctuations. The new experiments probe the physics of last scattering and are significantly narrowing the class of viable models. A new technique for measuring $\Omega_B$ is being developed utilizing the power-spectral signature of the temperature fluctuations, that already represents a triumph for dark matter dominated cosmological models with late reionization. The joint use of information from $z \simeq 1000$ (i.e. CMB) and from $z \simeq 0$ (i.e. LSS) will be a powerful tool for cosmology. The determination of the specific cosmological model that describes our Universe is an exciting challenge that still lies ahead of us, but the field is already providing answers to some fundamental questions.

## Acknowledgements

We would like to thank John Peacock for providing us with the LSS data, and Naoshi Sugiyama, Wayne Hu and Ted Bunn for many useful conversations. This work was supported in part by grants from the NSF.

## References and Notes

Acronyms in square brackets indicate experimental papers from which data have been taken.

1. White, M., Scott, D. & Silk, J. *Ann. Rev. Astron. & Astrophys.* **32**, 319 (1994).
2. Smoot, G. *et al. Astrophys. J.* **396**, L1 (1992).
3. Silk, J. *Astrophys. J.* **151**, 459 (1968).
4. Doroshkevich, A. G., Zel'dovich, Ya. B. & Sunyaev, R. A. *Sov. Astron.* **22**, 523 (1978).
5. Jørgensen, H. E., Kotok, E., Nasel'skiĭ, P. D. & Novikov, I. D. *Astr. Astrophys.* **294**, 639 (1994).
6. Hu, W. & Sugiyama, N. *Phys. Rev. D* **51**, 2599 (1994).
7. Bond, J. R., Crittenden, R., Davis, R. L., Efstathiou, G. & Steinhardt, P. J. *Phys. Rev. Lett.* **72**, 13 (1994).
8. Seljak, U. *Astrophys. J.* **435**, L87 (1994).
9. As well as the usual (scalar) density perturbations, inflation can generate (tensor) ripples in the background space-time. If present, these 'gravity waves' cause additional fluctuations in the microwave background. In inflationary models one generally expects $C_\ell^T \ll C_\ell^S$ if $n$ is close to 1.
10. Górski, K. M. *et al. Astrophys. J.* **430**, L89 (1994). [COBE]
11. Bunn, E. F. & Sugiyama, N. *Astrophys. J.* (in press, 1994).
12. Bunn, E. F., Scott, D. & White, M. *Astrophys. J.* **441**, L9 (1994).
13. Peebles, P. J. E. & Yu, J. T. *Astrophys. J.* **162**, 815 (1970).
14. Wilson, M. L. & Silk, J. *Astrophys. J.* **243**, 14 (1981).
15. Bond, J. R. & Efstathiou, G. *Mon. Not. R. astr. Soc.* **226**, 655 (1987).
16. Sugiyama, N. & Gouda, N. *Prog. Theor. Phys.* **88**, 803 (1992).
17. Kamionkowski, M., Spergel, D. N. & Sugiyama, N. *Astrophys. J.* **426**, L57 (1994).
18. Scott, D. & White, M. in *CMB Anisotropies Two Years After COBE*, ed. L. Krauss (World Scientific, Sihgapore, 1994), 214-228.
19. White, M. & Srednicki, M. *Astrophys. J.* **443**, 6 (1994).
20. Bond, J. R. in *Proceedings of the IUCAA Dedication Ceremonies*, ed. T. Padmanabhan (New York, John Wiley & Sons, 1993), in press.
21. Bunn, E., White, M., Srednicki, M. & Scott, D. *Astrophys. J.* **429**, 1 (1994).
22. Tucker, G. S., Griffin, G. S., Nguyen, H. T. & Peterson, J. B. *Astrophys. J.* **419**, L45 (1993). [WD]
23. Readhead, A. C. S., Lawrence, C. R., Myers, S. T., Sargent, W. L. W., Hardebeck, H. E. & Moffet, A. T. *Astrophys. J.* **346**, 566 (1989). [OVRO]
24. Subrahmanyan, R., Ekers, R. D., Sinclair, M. & Silk, J. *Mon. Not. R. astr. Soc.* **263**, 416 (1993). [ATCA]
25. White, M. & Scott, D. in *CMB Anisotropies Two Years After COBE*, ed. L. Krauss (World Scientific, Sihgapore, 1994), 254-258.
26. Scott, D., Srednicki, M. & White, M. *Astrophys. J.* **421**, L5 (1994).
27. Since $A_{\rm peak}$ is defined in terms of the power spectrum ($Q_{\rm flat}$ squared), the corresponding enhancement at $\ell \sim 200$ on Fig. 1 is $(1 + A_{\rm peak})^{1/2}$ times the amplitude at low $\ell$.
28. Bartlett, J. G. & Stebbins, A. *Astrophys. J.* **371**, 8 (1991).
29. Coulson, D., Ferreira, P., Graham, P. & Turok, N. *Nature* **368**, 27 (1994).
30. Silk, J. *Nature* **215**, 1155 (1967).
31. Sachs, R. K. & Wolfe, A. M. *Astrophys. J.* **147**, 73 (1967).
32. Sugiyama, N. & Silk, J. *Phys. Rev. Lett.* **73**, 509 (1994).
33. Hu, W., Sugiyama, N. & Bunn, E. *Astrophys. J.* (in press, 1995).
34. Tegmark, M. & Silk, J. *Astrophys. J.* **243**, 529 (1994).
35. Knox, L., & Turner, M. *Phys. Rev. Lett.* **73**, 3347 (1994).
36. White, M., Krauss, L. & Silk, J. *Astrophys. J.* **418**, 535 (1993).
37. Peacock, J. A. & Dodds, S. J. *Mon. Not. R. astr. Soc.* **267**, 1020 (1994).
38. Lyth, D. H. & Stewart, E. D. *Phys. Lett. B* **252**, 336 (1990).
39. Ratra, B. & Peebles, P. J. E. *Astrophys. J.* **432**, L5 (1994).
40. In an open geometry there is ambiguity in defining wavenumber. Here $k$ is the eigenvalue of the Laplacian. The upturn is due to the choice of $k$ as the $x$ ordinate; the total power is of course finite.
41. Bartlett, J. G., Blanchard, A., Silk, J. & Turner, M. S. *Science* **267**, 980 (1994).
42. White, M., Gelmini, G. & Silk, J. *Phys. Rev. D* **51**, 2669 (1994).
43. Efstathiou, G., Bond, J. R. & White, S. D. M. *Mon. Not. R. astr. Soc.* **258**, 1P (1992).
44. We note in passing that there is a 25% correction to the $\Omega_0^{-1.54}$ scaling used in (*43*) due to the integrated Sachs-Wolfe effect, since the potentials are not constant at late times in a $\Lambda$-dominated model (*45,46*).
45. Kofman, L. & Starobinskiĭ, A. A. *Sov. Astron. Lett.* **11**, 271 (1985).
48. Krauss, L. M. & Kernan, P. *Phys. Lett. B* (in press, 1995).
46. White, M. & Bunn, E. F. *Astrophys. J.* (in press, 1995).
47. Kofman, L., Gnedin, N. Y. & Bahcall, N. A., *Astrophys. J.* **413**, 1 (1993).
49. Lyth, D. H. & Liddle, A. R. *Astrophys. Lett. & Commun.* (in press, 1994).
50. Hu, W., Scott, D. & Silk, J. *Phys. Rev. D* **49**, 648 (1994).
51. Ganga, K., Page, L., Cheng, E. & Meyer, S. *Astrophys. J.* **432**, L15 (1994). [FIRS]
52. Hancock, S. *et al. Nature* **367**, 333 (1994). [Ten.]
53. Gundersen, J. O. *et al. Astrophys. J.* **443**, L57 (1994). [SP94]
54. Netterfield, C.B., Jarosik, N. C., Page, L. A., Wilkinson, D., Wollack, E. *Astrophys. J.* (in press, 1994). [SK]
55. Dragovan, M. *et al. Astrophys. J.* **427**, L67 (1993). [Pyth.]
56. de Bernardis, P. *et al. Astrophys. J.* **422**, L33 (1994). [ARGO]
57. Piccirillo, L. & Calisse, P. *Astrophys. J.* **411**, 529 (1993). [IAB]
58. Alsop, D. C. *et al. Astrophys. J.* **395**, 317 (1992). [MAX-2]
59. Gundersen, J. O. *et al. Astrophys. J.* **413**, L1 (1993). [MAX-3]
60. Devlin, M. *et al. Astrophys. J.* **430**, L1 (1994). [MAX-4]
61. Meinhold, P. R. *et al. Astrophys. J.* **409**, L1 (1993). [MAX-3]
62. Clapp, A. C. *et al. Astrophys. J.* **433**, L57 (1994). [MAX-4]
63. Cheng, E. S. *et al. Astrophys. J.* **422**, L37 (1994). [MSAM]

# Figure Captions

**Fig. 1.** The amplitude of '$\Delta T$' fluctuations in each experiment, as a function of scale (multipole $\ell \sim \theta^{-1}$). $Q_{\text{flat}}$ is the best-fitting amplitude of a flat power spectrum, quoted at the quadrupole (see Eq. 2). The vertical error bars are $\pm 1\sigma$, while the horizontal lines represent the half-power ranges of the window functions. The data points are listed in Table 2, where the references to the experiments are also given. We have plotted the data for 5 of the 6 MAX results as one point, with the discrepant $\mu$ Peg point plotted separately. The MSAM experiment has two independent modes. There are also three smaller-scale upper limits plotted at the 95%CL. The general rise in the area around $\ell \simeq 200$ can be interpreted as evidence for an adiabatic peak in the radiation power spectrum.

**Fig. 4.** The matter power spectrum, $P(k)$, on a range of scales, as inferred from Large–Scale Structure and Cosmic Microwave Background data. The boxes are $\pm 1\sigma$ values of $P(k)$ inferred from CMB measurements on large and intermediate scales assuming: *top* CDM ($\Omega_0 = 1$, $h = 0.5$, $\Omega_B = 0.03$) or MDM ($\Omega_\nu = 0.3$); *middle* an open universe inflationary model ($\Omega_0 = 0.3$, $h = 0.8$, $\Omega_B h^2 = 0.0125$); *bottom* $\Lambda$CDM ($h = 0.8$, $\Omega_B h^2 = 0.0125$, $\Omega_\Lambda = 0.7$). The horizontal width of each box represents the range of scales to which the experiment is most sensitive. The LSS data are a compilation taken from (*37*).

**Fig. 5.** The matter power spectrum, $P(k)$, for more 'realistic' specific choices of the model parameters. The data points are as in Fig. 4. The models are : *top* CDM ($\Omega_0 = 1$, $h = 0.45$, $\Omega_B h^2 = 0.02$, tilted to $n = 0.9$); *middle* open ($\Omega_0 = 0.4$, $h = 0.7$, $\Omega_B h^2 = 0.02$, $n = 1$); *bottom* $\Lambda$CDM ($h = 0.8$, $\Omega_B h^2 = 0.02$, $\Omega_\Lambda = 0.7$, $n = 0.95$ with a gravity wave component). In general these models provide a better fit than the more 'standard' models of Fig. 4, illustrating the potential power in using CMB and LSS data together. Note that the overall amplitude of the LSS data is uncertain to perhaps 20%.

# Tables

**Table 1.** The primary sources of temperature fluctuations roughly in order of increasing importance with decreasing angular scale.

| | |
|---|---|
| ○ $\Delta T/T = V_\odot/c$ | dipole anisotropy, where $V_\odot$ is our motion relative to the radiation; |
| ○ $\Delta T/T = -\delta\phi$ | gravitational potential or Sachs-Wolfe fluctuations; |
| ○ $\Delta T/T = \frac{1}{3}\delta\rho/\rho$ | density perturbations, if the perturbations are 'adiabatic'*; |
| ○ $\Delta T/T = -\frac{1}{3}\delta S$ | entropy perturbations, if the perturbations are 'isocurvature'*; |
| ○ $\Delta T/T = v/c$ | Doppler shifts, when the photons were last scattered. |

* Adiabatic initial conditions, naturally generated during inflation, have $(\delta\rho/\rho)_{\rm rad} = (4/3)(\delta\rho/\rho)_{\rm mat}$, so that the entropy is constant. Isocurvature initial conditions have $\delta\rho_{\rm rad} = -\delta\rho_{\rm mat}$, so that there is no perturbation to the total energy density.

**Table 2.** Summary of angular scales and measured fluctuations for current experiments. The parameters $\ell_0$, $\ell_1$ and $\ell_2$ are the peak and the lower and upper half-peak points of the window function, respectively. $Q_{\rm flat}$ is the best-fit amplitude for a flat spectrum through the window function, quoted at the quadrupole scale. The error bars are $\pm 1\sigma$.

| Experiment | $\ell_0$ | $\ell_1$ | $\ell_2$ | $Q_{\rm flat}(\mu K)$ |
|---|---|---|---|---|
| COBE (10) | — | — | 18 | $19.9 \pm 1.6$ |
| FIRS (51) | — | — | 30 | $19 \pm 5$ |
| Ten. (52) | 20 | 13 | 30 | $26 \pm 6$ |
| SP94 (53) | 67 | 32 | 110 | $26 \pm 6$ |
| SK (54) | 69 | 42 | 100 | $29 \pm 6$ |
| Pyth. (55) | 73 | 50 | 107 | $37 \pm 12$ |
| ARGO (56) | 107 | 53 | 180 | $25 \pm 6$ |
| IAB (57) | 125 | 60 | 205 | $61 \pm 27$ |
| MAX–2 ($\gamma$ UMi) (58) | 158 | 78 | 263 | $74 \pm 31$ |
| MAX–3 ($\gamma$ UMi) (59) | 158 | 78 | 263 | $50 \pm 11$ |
| MAX–4 ($\gamma$ UMi) (60) | 158 | 78 | 263 | $48 \pm 11$ |
| MAX–3 ($\mu$ Peg) (61) | 158 | 78 | 263 | $19 \pm 8$ |
| MAX–4 ($\sigma$ Her) (62) | 158 | 78 | 263 | $39 \pm 8$ |
| MAX–4 ($\iota$ Dra) (62) | 158 | 78 | 263 | $39 \pm 11$ |
| MSAM2 (63) | 143 | 69 | 234 | $40 \pm 14$ |
| MSAM3 (63) | 249 | 152 | 362 | $39 \pm 12$ |

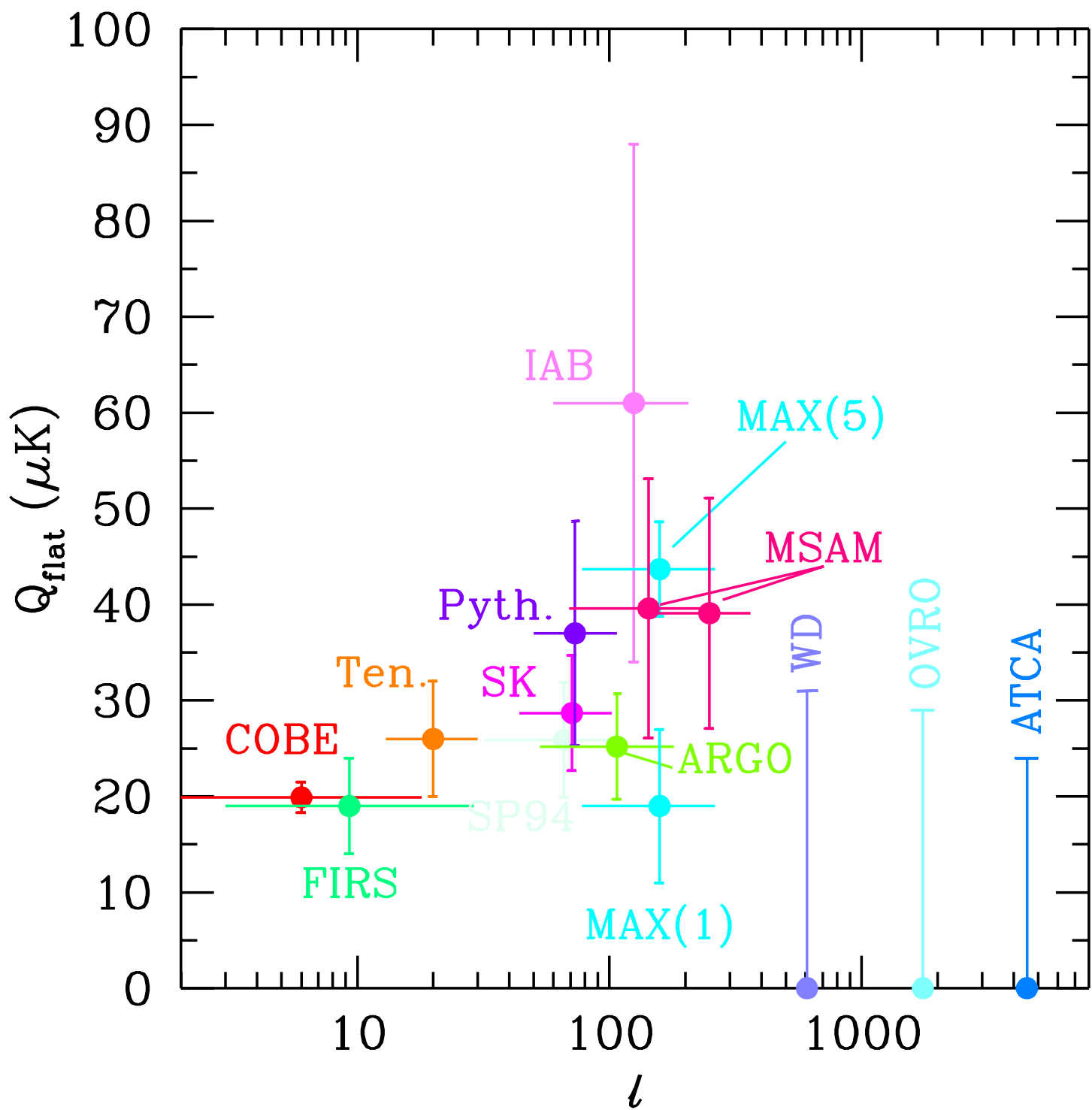

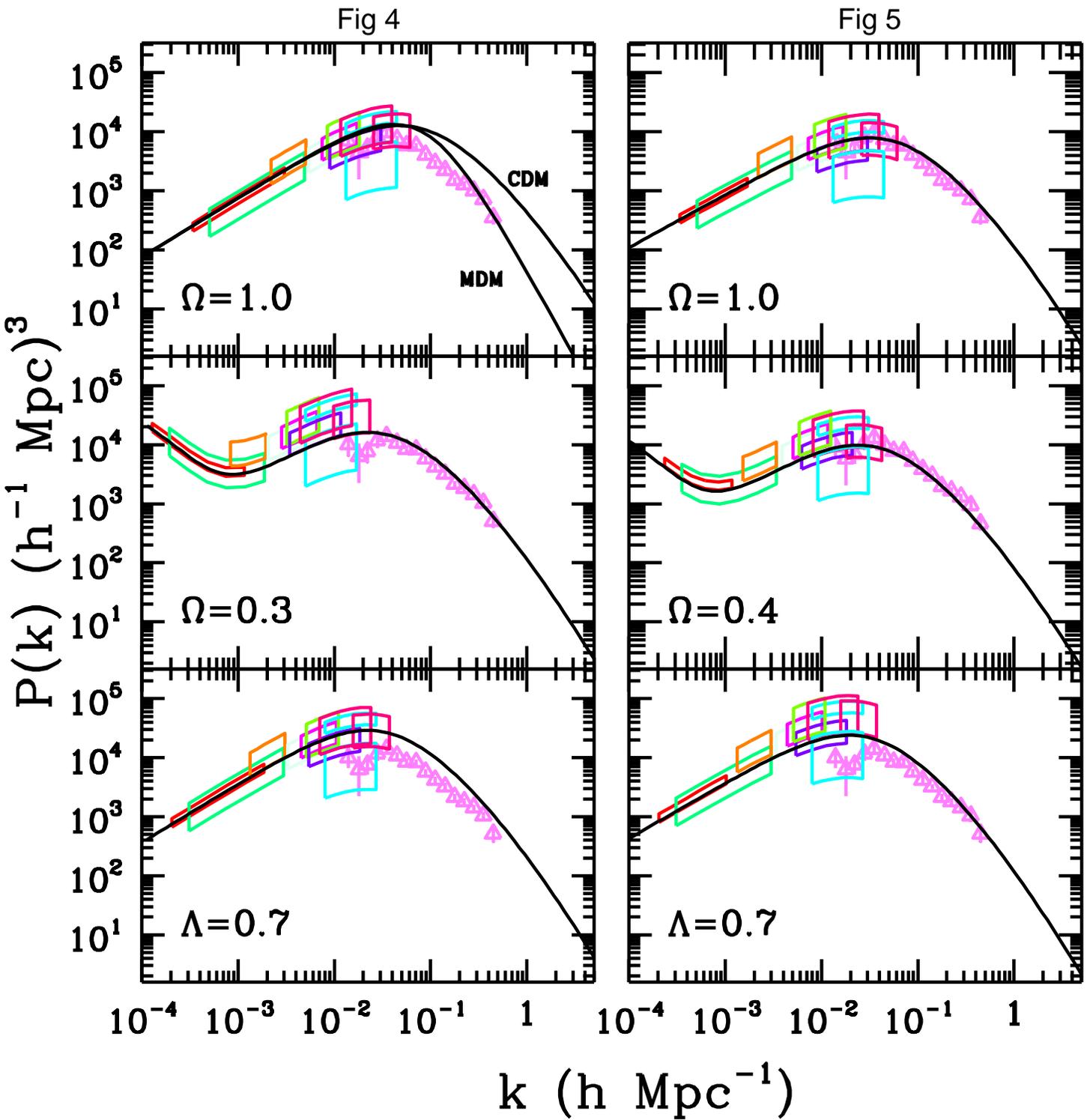